\documentclass[10pt,conference]{IEEEtran}
\usepackage{times}

\usepackage{amsmath}
\usepackage{amsfonts}
\usepackage{latexsym}
\usepackage{amssymb}
\usepackage{amsthm}
\usepackage{upref}
\usepackage{graphicx}
\usepackage{psfrag}

\usepackage{float}
\usepackage{color}
\usepackage[margin=0.75in]{geometry}
\usepackage{flushend}
\usepackage{multicol}

\newtheorem{thm}{Theorem$\!$}

\newtheorem{lem}[thm]{Lemma$\!$}

\newtheorem{prop}[thm]{Proposition$\!$}

\newtheorem{cor}[thm]{Corollary$\!$}

\newtheorem{defn}[thm]{Definition$\!$}

\newtheorem{xmpl}[thm]{Example$\!$}
\newenvironment{example}{\begin{xmpl}\hspace*{-1ex}{\bf}}{\end{xmpl}}

\newtheorem{cnstr}[thm]{Construction$\!$}

\newtheorem{algr}[thm]{Algorithm$\!$}

\newtheorem{conj}[thm]{Conjecture$\!$}

\begin{document}

\title{Representation-Oblivious Error Correction\\ by Natural Redundancy}

\author{
\IEEEauthorblockN{\textbf{Pulakesh Upadhyaya and Anxiao (Andrew) Jiang}}
\IEEEauthorblockA{Computer Science and Engineering Department, Texas A\&M University \\
{\it pulakesh@tamu.edu, ajiang@cse.tamu.edu}\vspace*{-4.0ex}} 
}

\maketitle

\begin{abstract}
Storage systems have a strong need for substantially improving their error correction capabilities, especially for long-term storage where the accumulating errors can exceed the decoding threshold of error-correcting codes (ECCs). In this work, a new scheme is presented that uses deep learning to perform soft decoding for noisy files based on their natural redundancy. The soft decoding result is then combined with ECCs for substantially better error correction performance. The scheme is representation-oblivious: it requires no prior knowledge on how data are represented (e.g., mapped from symbols to bits, compressed, and combined with meta data) in different types of files, which makes the solution more convenient to use for storage systems. Experimental results confirm that the scheme can substantially improve the ability to recover data for different types of files even when the bit error rates in the files have significantly exceeded the decoding threshold of the ECC. The code of this work has been publicly released.
\footnote{https://github.com/pulakeshupadhyaya/RACC\_NR}
\end{abstract}
        
\section{Introduction}
         
The amount of data in storage systems is increasing at an accelerating speed in the big data era.  Storage systems have a strong need for substantially improving their error correction capabilities, especially for long-term storage where the accumulating errors can exceed the decoding threshold of error-correcting codes (ECCs). Memory scrubbing alone is not a sufficient solution: even for nonvolatile memory systems such as SSDs (solid-state drives) that have fast read/write speeds,  scrubbing all data periodically is still too costly due to the volume of the data. Therefore it is highly necessary to find new techniques to assist ECCs and substantially enhance their error correction performance.

One promising technique is to use the internal redundancy in data for error correction, and combine it with ECC's decoding algorithm. This type of redundancy, called \emph{natural redundnacy}, has been explored in recent works~\cite{Li,Luo,Pulakesh2017Allerton,Wang2016Globecom}. In practical storage systems, many files are either uncompressed or compressed imperfectly, especially for languages and images because their highly complex data models make perfect compression infeasible due to prohibitively high computational complexities. The residual redundancy (i.e., natural redundnacy) in data can then be combined with the redundancy artificially added by ECCs (i.e., parity-check bits) for joint error correction. There is often plenty of natural redundancy in data. For instance, for the English language, state-of-the-art compression algorithms (e.g., syllable-based Burrows-Wheeler Transform) can compress it to 2 bits/character~\cite{Lansky}, which is still far from Shannon's estimation of 1.34 bits/character as the entropy of printed English~\cite{Shannon}. For images, their true entropy remains unknown. But recent progress in deep learning, such as the inpainting techniques for completing images~\cite{Xie}, suggests that the natural redundancy in even compressed images is still substantial. In~\cite{Pulakesh2017Allerton,Wang2016Globecom}, natural redundancy has been used to help ECCs -- including LDPC codes and polar codes -- correct errors and achieved significant performance improvement. Those works have addressed both languages and images, but mainly for texts in English compressed by Huffman codes or fixed-codebook LZW codes~\cite{Li}. 

In this paper, we study how to use natural redundancy for error correction in a more practical setting. We consider noisy file segments from files of different types (e.g., HTML, LaTeX, PDF and JPEG), and correct errors in them even when their bit error rates (BERs) have significantly exceeded the decoding threshold of the ECC. The scheme is \emph{representation-oblivious}: it requires no prior knowledge on how data are represented in those different file types, e.g., how symbols/characters are mapped to bits, how/whether data are compressed, and how meta data are used in those files. This approach makes the solution more convenient to use for storage systems. It is different from the previous works (e.g.,~\cite{Li,Pulakesh2017Allerton}), where the data (e.g., texts) are compressed by known compression algorithms (e.g., Huffman or LZW code) and without any additional data formatting (e.g., meta data or file formats) that brings more complexity. (In those works, the codebook of Huffman or LZW code is used for decoding. In this paper we do not use any codebook.) We take this representation-oblivious approach because in storage systems (such as SSDs), many file types have proprietary compression algorithms or file formats that are often unrevealed to the public, including to storage device manufacturers. Also, since error correction is a low-layer function in the storage architecture, the controllers of storage devices do not necessarily have access to file systems to get information on file types, data formats or compression algorithms. By taking the representation-oblivious approach, we can explore error correction schemes based on natural redundancy that are more widely usable in storage systems.

\begin{figure*}
\includegraphics[height=3.9cm, width=16cm]{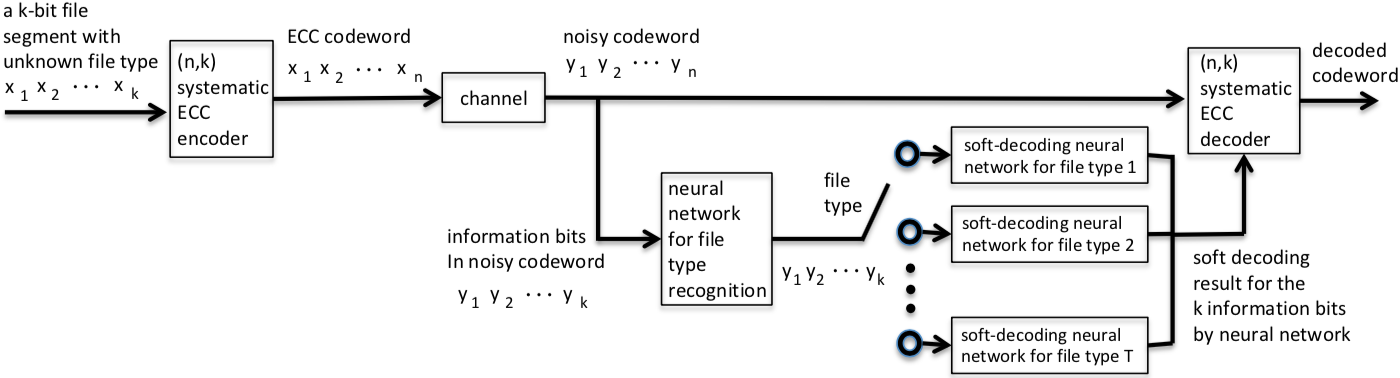}
\caption{Encoding and decoding scheme for a noisy file segment of an initially unknown file type. The $k$-bit file segment is encoded by a systematic $(n,k)$ ECC into an $n$-bit codeword. The codeword is transmitted through a channel to get a noisy codeword. Two neural networks use natural redundancy to decode the $k$ noisy information bits: the first network determines the file type of the file segment, and then a corresponding neural network for that file type performs soft decoding for the $k$ noisy information bits. The soft decoding result and the noisy codeword are both given to the ECC decoder for further error correction.}
\label{fig:codeModel}
\end{figure*}

The coding scheme of this paper is illustrated in Fig.~\ref{fig:codeModel}. 
When files are stored, each file is partitioned into segments of $k$ bits, and each file segment is encoded by a systematic $(n,k)$ ECC into a codeword of $n$ bits. Then each ECC codeword passes through a noisy channel, which models the errors in a storage device. During decoding, first, a deep neural network (DNN) uses the $k$ noisy information bits to recognize the file type (e.g. HTML, LaTeX, PDF or JPEG) of the file segment. Then, a second DNN for that file type performs soft decoding on the $k$ noisy information bits based on natural redundancy, and outputs $k$ probabilities, where for $i=1,2,\cdots,k$, the $i$-th output is the probability for the $i$-th information bit to be 1. The $k$ probabilities are given as additional information to the ECC's decoder. The ECC decoder then performs its decoding and outputs the final result. (In our experiments, the ECC is a systematic LDPC code, and the $k$ probabilities are combined with the initial LLRs (log-likelihood ratios) for information bits to obtain their updated LLRs. The LDPC code then runs its belief-propagation (BP) decoding algorithm.)

The above scheme can be extended to the case where the two decoders -- the decoder based on natural redundancy (NR decoder) and the ECC decoder -- perform iterative decoding between them.  That is, each decoder's output is given to the other decoder as input, and the decoding process iterates between the two decoders. Iterative decoding of this type for the English language compressed by known compression algorithms has been studied in~\cite{Luo,Pulakesh2017Allerton}. The scheme can also be extended to the case where an ECC codeword may contain multiple file segments of multiple file types. For simplicity, such extensions are not explored in this paper.

This work has several contributions. First, it designs a deep neural network that recognizes file types with high accuracy from noisy bits. For error correction, this DNN helps recognize the type of natural redundancy in the noisy data.

Second, it designs deep neural networks for decoding data with natural redundancy, where the data have errors from the binary-symmetric channel (BSC). The DNNs perform soft decoding instead of hard-decision decoding, which can be more useful for ECCs such as LDPC codes. Since the data used to train the DNNs do not contain soft decoding results, we design a new portfolio theory-based approach to train the DNNs. The results show that the DNNs can learn soft decoding with high accuracy, even though the training data are extremely sparse compared to possible data patterns.

Third, the paper presents a scheme that combines the natural redundancy based decoding, which applies deep learning to noisy file segments of different file types, with ECC decoding. The experimental results confirm that the scheme substantially improves the reliability of different types of files.

There have been numerous recent works on using deep learning for information theory, especially for wireless and optical communications. They mainly focus on using deep learning to model complex channels, design codes, and approximate or improve decoding algorithms~\cite{HKim2,Nachmani2}. 
In contrast to those works, this paper focuses on using deep learning for \emph{data} with complex structures, and explore error correction for such complex data. These different directions can complement each other in a communication or storage system with both complex data and complex channels.

The rest of the paper is organized as follows. In Section II, we develop deep neural networks for recognizing file types from noisy bits. In Section III, we present a portfolio theory-based method that teaches a deep neural network soft decoding. We then design deep neural networks that perform soft decoding on file segments. In Section IV, we combine the natural redundancy (NR) decoder with an LDPC decoder for substantially enhanced error correction performance. In Section V, we present concluding remarks.

\section{File Type Recognition using Deep Learning} 

In this section, we present a \emph{Deep Neural Network} (DNN) for file type recognition. The DNN takes a noisy file segment of $k$ bits, $(y_{1},y_{2},\cdots,y_{k})$, as input, and outputs one of $T$ file types (e.g., HTML, LaTeX, PDF or JPEG). The errors in the file segment come from a binary-symmetric channel (BSC) of bit-error rate (BER) $p$. We first introduce the architecture of the DNN and its training method. We then present the experimental results, which show that it achieves high accuracy for file type recognition.

\subsection{DNN Architecture and Training}

\begin{figure}
\includegraphics[height=6cm, width=8cm]{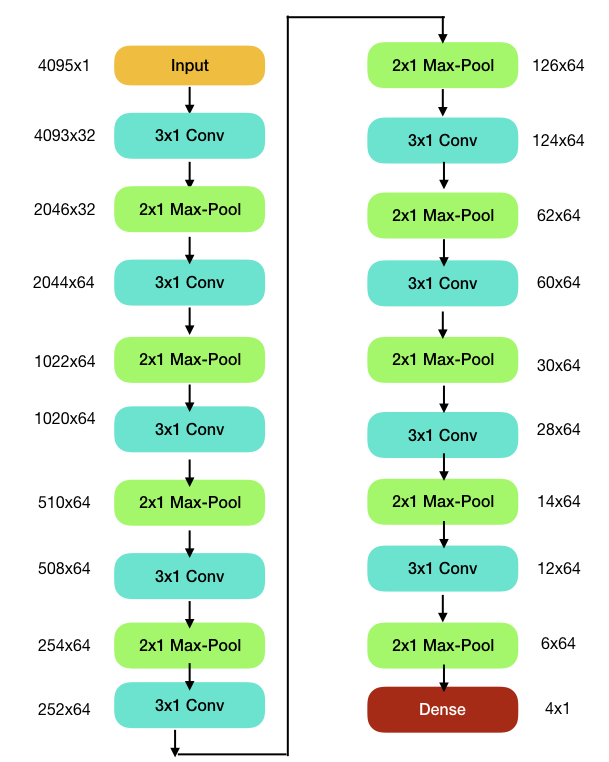}
\caption{ Architecture of the CNN (convolutional neural network) for File Type Recognition. Its input is a noisy file segment of 4095 bits, and its output corresponds to 4 candidate file types (HTML, LaTex, PDF and JPEG). The numbers beside each layer (namely, $4095\times1$, $4093 \times 32$, $\cdots$, $4\times 1$) are the dimension sizes of the layer's output data. The numbers inside each layer (namely, $3\times 1$ or $2 \times 1$) are the dimension sizes of the corresponding feature-map filter or pooling window.}
\label{fig:FTR}
\end{figure}

Our DNN architecture is shown in Fig.~\ref{fig:FTR}. It is a Convolutional Neural Network (CNN) that takes the $k$ bits of a noisy file segment as input. In our experiments, we let $k=4095$. (The LDPC code we use is a $(4376,4095)$ code designed by MacKay~\cite{Mackay}, which can tolerate BER of 0.2\%. Both the code length and the BER are in the typical range of parameters for storage systems.) The CNN has $T$ outputs that correspond to the $T$ possible file types, namely, the $T$ classification results. The output with the highest value leads to the selection of the corresponding file type. In our experiments, we consider four file types: HTML, LaTeX, PDF and JPEG. So $T=4$. Note that HTML and LaTeX files are both text sequences but have different file structures; PDF files contain both texts and images; and JPEG files are images. In the following, we will present DNNs and experiments using those parameters for the convenience of presentation. Note that the designs can be extended to other file-segment lengths and more file types. 

In Fig.~\ref{fig:FTR}, there are $L=9$ convolution layers \{$C_1,C_2,...C_L\}$ where each layer $C_d$ (for $d=1,2,\cdots,L$) is followed by a max pooling layer $M_d$. The last max pooling layer $M_L$ is followed by a dense layer $D$. 
For $d=1,2,\cdots,L$, let $n_d$ denote the number of feature maps of the convolution layer $C_d$. (In Fig.~\ref{fig:FTR}, $n_{1}=32$ and $n_{2}=n_{3}=\cdots = n_{9} = 64$.) Those feature maps are obtained by taking convolution on the output of the previous layer using $n_d$ filters of size $l_d = 3$. In its subsequent max pooling layer $M_d$, pooling windows of size $2$ are applied to each feature map of $C_d$ with a stride of two. Let $K_d$ denote the length of each feature map (in the dimension of the CNN's input) of the layer $C_d$. Then $K_{1} = k - l_{d} +1$, and $K_d = \lfloor{\frac{K_{d-1}}{2}} \rfloor - l_d + 1$ for $2 \le d \le L$. 




The CNN uses \emph{ReLU} and \emph{sigmoid} as the activation function of its convolutional layers and output layer, respectively. It uses \textit{cross entropy} as its loss function. Its optimizer is chosen to be an \textit{Ada Delta Optimizer}, whose parameters are: learning rate = 1.0, $\rho$ = 0.95, $\epsilon$ = \emph{none} and decay = 0. During training, each mini-batch has 100 training samples, where each training sample consists of a noisy file segment and its file-type label (i.e., one of the $T$ file types). 

A large dataset has been used to train and test the CNN. For each of the $T=4$ file types, 24,000 noiseless file segments are used for training data, 4,000 noiseless file segments are used for validation data, and 4,800 noiseless file segments are used for test data. During training and testing, random errors of BER $p$ are added to each file segment, where each file segment uses an independently generated error pattern. 

\subsection{Experimental Performance}


\begin{table} 
\centering
\caption{Bit error rate (BER) vs Test Accuracy for File Type Recognition (FTR). Here the ``overall test accuracy" is for all 4 types of files together. The last four columns show the test accuracy for each individual type of files. (Their average value is the overall test accuracy.)}
\label{table1}
\begin{tabular}{|l|l|l|l|l|l|}
\hline 
Bit Error& Overall & HTML& JPEG & PDF  & LaTeX \\
Rate & Test & Test & Test & Test & Test \\
(BER) & Accuracy & Accuracy  & Accuracy & Accuracy & Accuracy \\ 
\hline 
0.2\% & 99.61\%  & 99.98\% & 99.52\% & 99.17\% & 99.77\% \\ 
0.4\% & 99.69\%  & 99.96\% & 99.60\% & 99.25\% & 99.96\% \\ 
0.6\% & 99.60\%  & 99.94\% & 99.48\% & 99.06\% & 99.90\% \\ 
0.8\% & 99.69\%  & 99.98\% & 99.50\% & 99.35\% & 99.92\% \\ 
1.2\% & 99.66\% & 99.96\% & 99.23\% &  99.48\% & 99.96\%  \\
1.6\% & 99.58\% & 99.96\% & 99.60\% &  98.83\% & 99.92\% \\
\hline 
\end{tabular}
\end{table}


The $(4376,4095)$ LDPC code used in our experiments can correct errors of BER up to 0.2\% \emph{by itself}. (That is, when it is used in the conventional way without the extra help of natural redundancy, it has a decoding threshold of 0.2\%.) Our goal is to use the natural redundancy in file segments to correct errors of substantially higher BERs. So we have selected the target BER $p$ with substantially higher values, ranging from $0.2\%$ to $1.6\%$. We then train the CNN with the given target BER $p$.

We measure the performance of the CNN by the \emph{accuracy} of file type recognition (FTR), which is defined as the fraction of file segments whose file types are recognized correctly. 
The CNN is trained using the training and validation data. Its final performance is measured using the test data, where file segments of the $T=4$ file types are randomly mixed. The test performance is shown in Table~\ref{table1}. It can be seen that file types can be recognized by the CNN with high accuracy: for all BERs, the accuracy is close to 1. 



We can also examine the accuracy for recognizing each file type, and see if there is variance in performance from file type to file type. The results are shown in the last four columns of Table ~\ref{table1}. It can be seen that overall, the accuracy is constantly high for all file types.

The CNN's performance compares favorably with existing results on FTR, which has been studied previously for applications such as disk recovery. The work~\cite{Calhoun} considered a classification method for a pair of file types using Fisher's linear discriminant and longest common subsequence methods. The accuracy ranges between $87\%$ and $99\%$ depending on which pair of file types are considered. The work~\cite{Fitzgerald} introduced an NLP (natural language processing) based method, where unigram and bigram counts of bytes and other statistics are used to generate feature representation, which is then followed by support vector machine (SVM) for classification of various file types. The classification accuracy varies from $17.4\%$ for JPEG files, $62.5 \%$ for PDF files to $94.8\%$ for HTML files. The work~\cite{Amirani} used PCA (principal component analysis) and a feed-forward auto-associative unsupervised neural network for feature extraction, and a three layer multi-layer perceptron network for classification. The classification accuracy is  $98.33\%$ for six file types while considering entire files instead of file segments. Our deep-learning based method can be seen to achieve high performance, without the need to train separate modules for feature extraction and classification. 

The CNN has robust performance because it works well not only for the BER it is trained for, but also for other BERs in the considered range. (For example, a CNN trained for $BER=1.2\%$ also works well for other BERs in the range $[0.2\%,2.0\%]$.) For succinctness we skip the details. The robustness of the overall error correction performance for different BERs will be presented in Section IV.

\section{Soft Decoding by Deep Neural Networks}

In this section, we study how to design deep neural networks that can perform soft decoding on noisy file segments. For each of the $T$ file types, we will design and train a different DNN, because different types of files have different types of natural redundancy. Given a file type, we will design a DNN whose input is a noisy file segment of $k$ bits 
$Y = (y_{1},y_{2},\cdots,y_{k})$. As before, the errors in the noisy file segment come from a binary-symmetric channel (BSC) of bit-error rate (BER) $p$. The output of the DNN is a vector $Q=(q_{1},q_{2},\cdots,q_{k})$, where for $i=1,2,\cdots,k$, the real-valued output $q_{i} \in [0,1]$ represents the DNN's belief that for the $i$-th bit in the file segment, the probability that its correct value should be 1 is $q_{i}$. In other words, if we use 
$X = (x_{1},x_{2},\cdots,x_{k})$ 
to denote an error-free file segment, and let it pass through a BSC of BER $p$ to obtain a noisy file segment 
$Y = (y_{1},y_{2},\cdots,y_{k})$, then $q_{i}$ is the DNN's estimation for 
$Pr\{x_{i}=1~|~Y,p\}$. Note that the $k$ bits are not independent of each other because of the natural redundancy in them. So 
$Pr\{x_{i}=1~|~Y,p\}$ depends  on not only $y_i$ and $p$, but also the overall value of 
$Y$. The goal of the DNN is to learn the natural redundancy in file segments, and use it to make the probability estimation $q_{i}$ be as close to 
the true probability $Pr\{x_{i}=1~|~Y,p\}$ 
as possible, for each $i$ and for each possible value 
$Y$ of the noisy file segment.

Several notable challenges exist. First, there are no training data of the \emph{form} needed by the deep neural network (DNN) for soft decoding. Every sample of the training data is of the form 
$(Y,X)$, namely a noisy file segment and its correct value. However, the binary vector 
$X$ 
is not the wanted output of the DNN; instead, the DNN needs to output a vector of probabilities. 
Second, the true distribution 
$Pr\{x_{i}~|~Y,p\}$ 
is unknown during training. The distribution depends on both BER $p$ and natural redundancy, and the latter is what the neural network needs to learn, so it is not given \emph{a prior}. Yet we want the DNN's output to converge to that distribution. And due to the sparsity of training data compared to the extremely large sample space of noisy file segments (which has $2^{k} = 2^{4095}$ possible values), it is practically infeasible to learn the true distribution based on statistical counting. 

Note that in many applications of deep learning (e.g., object recognition), DNN layers using \textit{sigmoid} or \textit{softmax} functions are often seen as outputting probabilities. However, those outputs are only variables that mimic probabilities to some very limited extent, and their accuracy is far from sufficient for the task of error correction. So a new technique is needed.

In the following, we present a new approach based on portfolio theory that teaches a DNN \emph{soft decoding}. We first present the idea, and verify its performance for data with small samples spaces. We then extend it to file segments, which have complex natural redundancy and a very large sample space.

\subsection{Portfolio Theory based Soft Decoding}

Consider a channel, whose input is a variable $X \in \{a_{i}~|~1 \le i \le K\}$, and whose output is a single bit $b \in \{0,1\}$. For $i=1,2,\cdots, K$, let $p_{i} \triangleq Pr\{b = 1 ~|~ X = a_{i}\}$ be the probability that the channel's output is 1 given that its input is $a_{i}$. Consider a sequence of $N$ such input-output pairs of the channel \[(X_{1},b_{1}),~(X_{2},b_{2}),~\cdots,~(X_{N},b_{N})\] where for $j=1,2,\cdots,N$, $X_{j}$ and $b_{j}$ are the input and output of the channel, respectively, for the $j$-th use of the channel. Now assume that we do not know the channel's transition probabilities $p_{1}$, $p_2$, $\cdots$, $p_{K}$.
Instead, we have the sequence of $N$ input-output pairs, and want to use them to estimate those transition probabilities. (We would like to achieve this goal without counting how many times the channel output is 1 for every given channel input value, because when this method is applied to file segments later, the channel will have $K=2^{k} = 2^{4095}$ possible input values, which is too large for counting to work due to the sparsity of training data and the memory constraint.)

Now suppose that we have derived a policy, which estimates the probability $Pr\{b = 1 ~|~ X = a_{i}\}$ as $q_{i}$ (when its true value should be $p_{i}$). Consider the following game on horse race, which is between two horses -- a white horse and a black horse -- and takes $X$ as its environment parameter (e.g., the wind speed, temperature, etc. at the race). Let $p_{i}$ (respectively, $1-p_{i}$) denote the probability that the white (respectively, black) horse wins when $X = a_{i}$. We bet on a sequence of $N$ races. For $j=1,2,\cdots,N$, for the $j$-th race, if its environment parameter $X = a_{i}$ (for some $i \in \{1,2,\cdots,K\}$), we bet a fraction of $q_{i}$ of our money on the white horse, and bet the remaining $1-q_{i}$ of our money on the black horse. If the white (respectively, black) horse wins in that race -- which corresponds to the channel output $b=1$ (respectively, $b=0$) -- the money we get is the amount of money we bet on the white (respectively, black) horse times some constant $c$. Now let us define a variable $S_j$ for the $j$-th race: if the white horse wins, we let $S_j = q_i$; otherwise, we let $S_j = 1-q_i$. 

Suppose that we started with 1 dollar. After the $N$ races, the money we have is $c^{N}\prod\limits_{j=1}^{N} S_j$. Define the \emph{doubling rate} as $$R = \frac{{log_2(
\prod\limits_{j=1}^{N} S_j
)}}{N} = \frac{1}{N}\sum\limits_{j=1}^{N} log(S_j).$$ 


\begin{example}
Let $K = 4$, and $N = 5$. Assume that we get the following sequence:
$(X_1,b_1) = (a_2,0),(X_2,b_2) = (a_1,1),(X_3,b_3) = (a_3,1),(X_4,b_4) = (a_3,0), (X_5,b_5) = (a_4,0)$. Then the doubling rate is
$R = \frac{1}{5} [log_2(1-q_2) + log_{2}(q_1)+log_{2}(q_3)+log_2(1-q_3)+log_2(1-q_4)] $.
\end{example}

By portfolio theory~\cite{Cover}, when $N\to \infty$, the doubling rate $R$ is maximized only if $q_i = p_i$ for $i=1,2,\cdots,K$. Therefore, to learn the transition probabilities  $p_{1}$, $p_2$, $\cdots$, $p_{K}$, we can design a neural network (NN) that takes the sequence of $N$ channel input-output pairs \[(X_{1},b_{1}),~(X_{2},b_{2}),~\cdots,~(X_{N},b_{N})\] as training data, and define the loss function of the NN as $-R$ namely, the \emph{negative doubling rate}. For $j=1,2,\cdots,N$, each $X_j$ is an input to the NN, and -- assuming $X_j = a_i$ for some $i$ -- the NN's output is considered to be $q_i$; and based on whether $b_j$ is 1 or 0, an additive term of $\log_{2}q_i$ or $\log_{2}(1-q_i)$ is included in the loss function. As the NN is trained, it \emph{minimizes} the loss function, which is equivalent to learning the correct transition probabilities and maximizing the doubling rate $R$.

In practice, the NN needs to gradually train its weights as it gets more and more training data, and $N$ cannot be infinite. So we need to partition the channel input-output pairs into batches, and let the NN use every batch to compute its loss functions and adjust its weights. To verify if the NN can learn the true transition probabilities effectively using batches of small sizes, we use the following experiments. 

For $K \ge 2$, we design a NN as follows. We take $N = K \times 50000$ channel input-output pairs in total. (Here we select each transition probability $p_{i}$ uniformly randomly from the range $(0,1)$. Then, given $p_1,p_2,\cdots,p_K$, we generate the $N$ channel input-output samples following those transition probabilities.) Let the batch size be $K \times 50$. Let the NN have three layers: an input layer, a hidden layer, and an output layer. The input layer uses one-hot encoding for the $K$ possible values for $X$; so the size of the input layer is  $K \times 1$. The size of the hidden layer is set to $K \times 1$ and is fully connected to the input layer. The size of the output layer is $1 \times 1$ (namely, just one number).
(For illustration, the architectures of the NNs for $K = 2$ and $K = 200$ are shown in Fig. 2.) 
After the NN is trained, for each input $X=a_i$, it can output the corresponding probability estimation $q_i$ (for $i=1,2,\cdots,K$).

\begin{figure}
\label{fig:PT1}
\includegraphics[height=3cm,width=8cm]{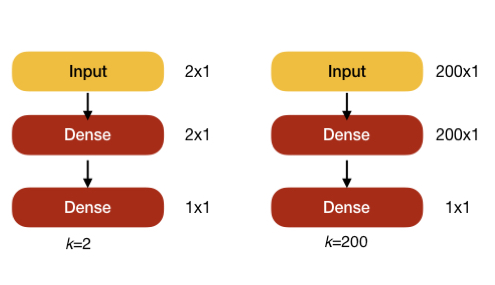}
\caption{Neural networks  for $K = 2$ (left) and $K=200$ (right).}
\end{figure}

We measure the distance between the true channel transition probabilities and the NN's estimation by their average Kullback-Liebler (KL) divergence
$$\Delta_K = \frac{1}{K}\sum\limits_{i=1}^{K} D(p_i \mid \mid q_i),$$ where $$ D(p_i \mid \mid q_i) = p_i log_2 \frac{p_i}{q_i} + (1-p_i)log_2 \frac{1-p_i}{1-q_i}.$$ The average KL divergence for different values of $K$ are shown in Table~\ref{table4}. It can be seen that the KL divergence is very small, which means that the NN has learned the true transition probabilities well.

\begin{table}
\centering 
\caption{Average KL Divergence between true and learned transition probabilities}
\label{table4} 
\begin{tabular}{|c|c|c|c|c|c|c|}
\hline 
$K$ & 2 & 4 & 10 &100 & 200  \\ 
\hline
$\Delta_K$ & 0.000069 & 0.000022 & 0.00015 & 0.00023 & 0.00021\\
 \hline 
\end{tabular}
\end{table}

\subsection{Soft Decoding for Noisy File Segments}

In the previous subsection, we have shown that the portfolio theory-based approach, which sets the NN's loss function as the negative doubling rate, works well for relatively simple channel models. However, when we apply this approach to file segments, several challenges appear. First, the output is no longer the probability for only one bit $b$; instead, it consists of $k$ probabilities $q_1,q_2,\cdots,q_k$ for the $k$ bits in the file segment. Our DNN needs to estimate them jointly using one network architecture. Second, in the experiments of the last subsection, the $K$ transition probabilities $p_1,p_2,\cdots,p_K$ are chosen independently and therefore have a simple structure; however, for file segments, the natural redundancy can be very complex, which can make the channel's transition probabilities be highly correlated and exhibit complex structures. Third, for file segments, the DNN's input can theoretically take $K=2^k = 2^{4095}$ possible values, which is a huge number and makes the training data very sparse. So it is not simple to see whether the DNN can learn the transition probabilities well given the sparsity of training data. 

In this subsection, we design DNNs for the soft decoding of noisy file segments. Our DNN architecture is related to auto-encoders. It consists of convolution layers followed by deconvolution layers. (Deconvolution layers may be seen as reverse operations of convolution layers. Interested readers can refer to~\cite{Chollet} for more details.) Auto-encoders are good choices for various applications related to denoising~\cite{Long,Vincent}.

\begin{figure*} 
\includegraphics[height=6.1cm, width=16cm]{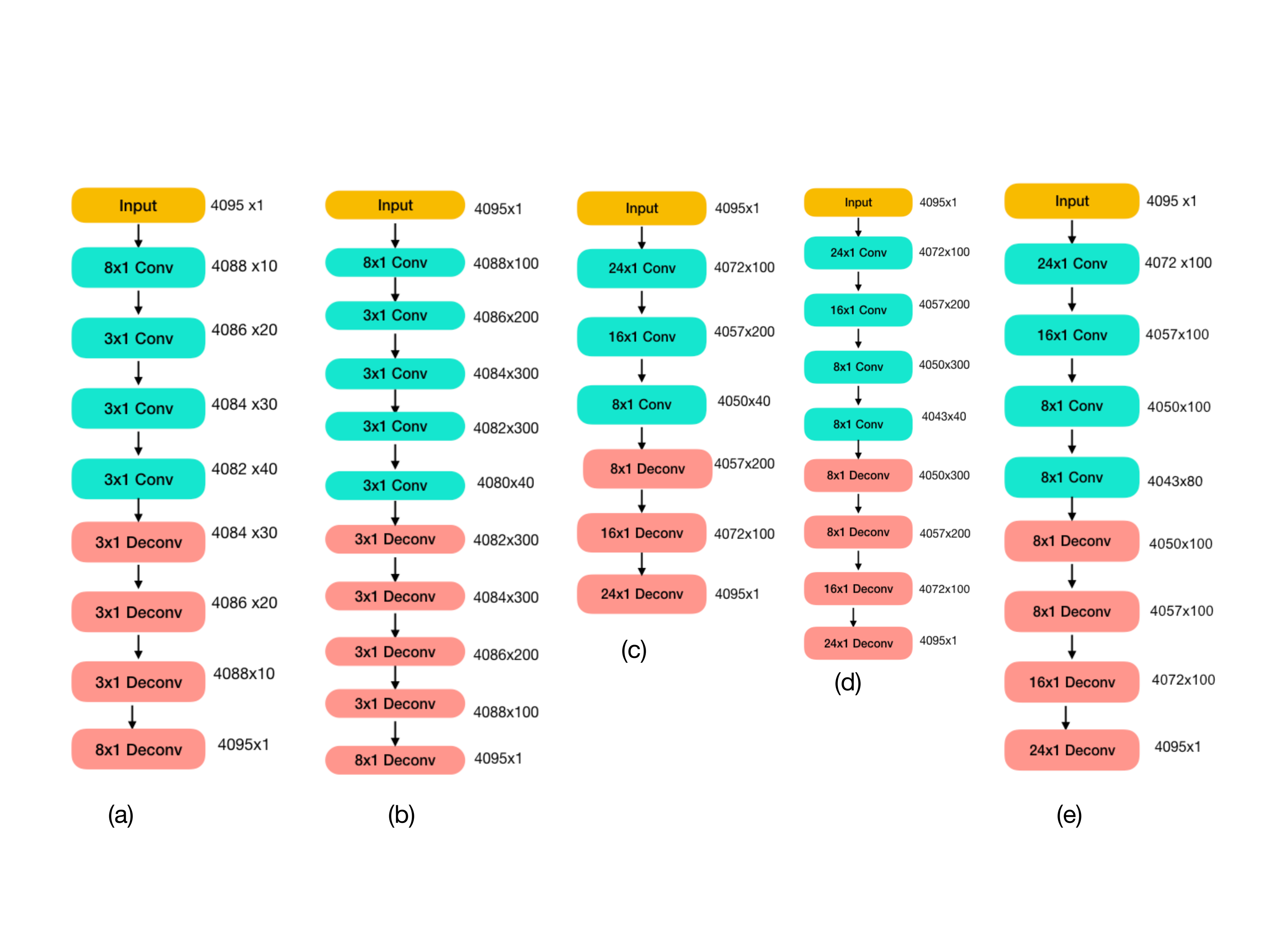}
\caption{Architectures of deep neural networks (DNNs) for soft decoding of noisy file segments. (a) DNN architecture for HTML files for $p = 0.8\%$, 1.2\%, 1.6\%, (b) DNN architecture for LaTex  files for $p = 0.8\%$, 1.2\%, 1.6\%, (c) DNN architecture for PDF and JPEG files when $p = 0.8\%$,  (d) DNN architecture for PDF files when $p = 1.2\%$, 1.6\%, (e) DNN architecture for JPEG files when $p = 1.2\%$, 1.6\%.}
\label{fig:ConvDeconv}

\includegraphics[height=6.7cm, width=16cm]{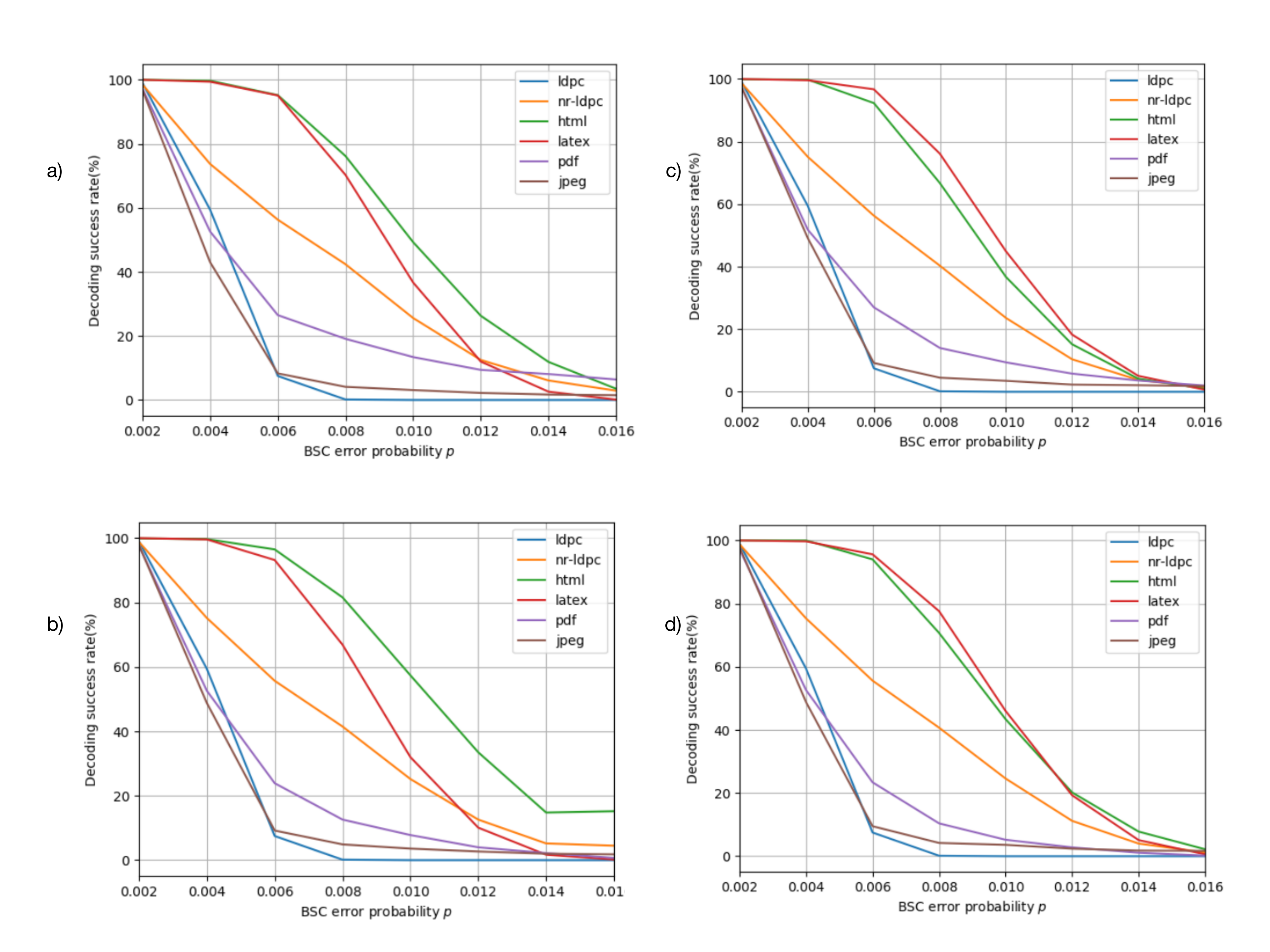}
\caption{ Decoding success rate vs bit error rate for (a) $p_{DNN}$ = $1.0\%$ , (b) $p_{DNN}$ = $1.2\%$, (c) $p_{DNN}$ = $1.4\%,$ (d) $p_{DNN}$ = $1.6\%$ } 
\label{fig:PERF}

\end{figure*}

 We illustrate our DNN architecture through some examples. Let $p$ be the BER of the binary-symmetric channel that adds errors to file segments. Consider $p = 0.8\%$, 1.2\%, 1.6\%. The DNN architecture for HTML (respectively, LaTeX) files, for all the above values of $p$, is shown in Fig.~\ref{fig:ConvDeconv} (a) (respectively, in Fig.~\ref{fig:ConvDeconv} (b)). When $p = 0.8\%$, the DNN architecture for both PDF and JPEG files is shown in Fig.~\ref{fig:ConvDeconv} (c). When  $p = 1.2\%$, 1.6\%, the DNN architecture for PDF (respectively, JPEG) files is shown in Fig. ~\ref{fig:ConvDeconv} (d) (respectively, in Fig.~\ref{fig:ConvDeconv} (e).

Since the transition probabilities corresponding to file segments are unknown, we cannot measure the performance of the DNN directly using the KL divergence. However, the experiments for error correction, which are to be presented in the next section, will confirm that the soft decoding result from the DNN is very useful for error correction.

\section{Error Correction for Noisy File Segments}

In this section, we combine the soft decoding output of the DNN -- which was presented in the previous section -- with an LDPC code for enhanced error correction performance. We adopt a \emph{robust scheme} here: the DNNs for file-type recognition and for soft decoding have been trained with a constant BER $p_{DNN}$, but they are used for a wide range of BERs $p$ for the BSC channel. (For example, the DNNs may be trained just for $p_{DNN} = 1.2\%$, but are used for any BER $p$ from 0.2\% to 1.6\% in experiments here.) We choose this robust scheme because when DNNs are designed, the future BER in data can be highly unpredictable. That is exactly why errors may exceed ECC's thresholds for long-term storage, and why natural redundancy can become useful for error correction.

Given a noisy systematic LDPC codeword, we first use a DNN to recognize its file type based on its $k$ noisy information bits. Then a second DNN for that file type is used to do soft decoding for the $k$ noisy information bits, and output $k$ probabilities: for $i=1,2,\cdots,k$, the $i$-th output $p_i$ represents the estimated probability for the $i$-th information bit to be 1. Those $k$ probabilities can be readily turned into LLRs (log-likelihood ratios) for the information bits using the formula $LLR_{i}^{DNN} = \log (\frac{1-p_{i}}{p_{i}})$. For $i=1,2,\cdots,n$, let $LLR_{i}^{channel}$ be the LLR for the $i$-th codeword bit (with $1 \le i \le k$ for information bits, and $k+1 \le i \le n$ for parity-check bits) derived for the binary-symmetric channel, which is either $\log(\frac{1-p}{p})$ (if the received codeword bit is 0) or $\log(\frac{p}{1-p})$ (if the received codeword bit is 1). Then we let the \emph{initial LLR} for the $i$-th codeword bit be $$LLR_{i}^{int} = LLR_{i}^{channel} + LLR_{i}^{DNN}$$ for $1\le i\le k$, and $LLR_{i}^{int} = LLR_{i}^{channel}$ for $k+1\le i\le n$. 
We then do belief-propagation (BP) decoding using the initial LLRs, and get the final result.

Note that there is a positive -- although very small -- chance that the file type will be recognized incorrectly. In that case, the incorrect soft-decoding DNN will be used. And that is accounted for in the overall decoding performance. 

We measure the performance of the error correction scheme by the percentage of codewords that are decoded correctly, which we call \emph{Decoding Success Rate}. (Let us call the scheme the \emph{NR-LDPC decoder}, since it combines decoding based on natural redundancy and the LDPC code.) We focus on BERs that are beyond the decoding threshold of the LDPC code, because natural redundancy becomes helpful in such cases. Note that the $(4376,4095)$ LDPC code used in our experiments has a decoding threshold of $BER=0.2\%$. In our experiments, we focus on BERs $p$ that are not only beyond the decoding threshold, but also can be significantly larger: $p \in [0.2\%,~1.6\%]$.

The experimental results for $p_{DNN} = 1.0\%$ is presented in Fig.~\ref{fig:PERF} (a). Here the $x$-axis is the channel error probability $p$, and the $y$-axis is the decoding success rate. (For each $p$, 1000 file segments with independent random error patterns have been used in experiments.) The curve for ``ldpc'' is the performance of the LDPC decoder alone, and the curve for ``nr-lpdc'' is for the NR-LDPC decoder. It can be seen that the NR-LDPC decoder achieves significantly higher performance. For example, as $p=0.6\%$, the decoding success rate of the NR-LDPC decoder is approximately 4 times as high as the LDPC decoder.

The figure also shows the performance for each of the 4 file types. (The 4 curves are labelled by html, latex, pdf, jpeg, respectively. Their average value becomes the curve for ``nr--ldpc''.) It shows that the error correction performance for HTML and LaTex files are significantly better than for PDF and JPEG files. It is probably because the former two mainly consist of languages, for which the soft-decoding DNNs are better at  finding their patterns and mining their natural redundancy, while PDF is a mixture of languages and images and JPEG is image only. It is interesting to notice that even for JPEG files, when $p > 0.6\%$, the NR-LDPC decoder again performs better than the LDPC decoder, which means the DNNs can extract natural redundancy from images, too. Fig.~\ref{fig:PERF} (b) to Fig.~\ref{fig:PERF} (d) show the performance for $p_{DNN}$ = 1.2\%, 1.4\% and 1.6\%, respectively. The NR-LDPC decoder performs equally well in those cases, which proves the value of natural redundancy for decoding.

\section{Conclusion}

This paper presents a new scheme that combines natural redundancy with LDPC codes for error correction. It is applied to noisy file segments of initially unknown file types, -- which is the first of its kind to the best of our knowledge, --- and shows substantial performance improvement compared to the original LDPC decoding scheme. The study can be extended to more types of natural redundancy in various types of files, more DNN architectures, and more ways to combine the NR decoder and ECC decoder, such as through iterative decoding between the two. Those remain as our future research directions.

\end{document}